\documentstyle[preprint,aps,epsf]{revtex}
\def\rts{\sqrt s}
\def\wp{W^+}
\def\wm{W^-}
\def\anti{\overline}
\def\fbi{~{\rm fb}^{-1}}

\def\mev{~{\rm MeV}}
\def\gev{~{\rm GeV}}

\def\anti{\overline}

\def\eg{{\it e.g.}}

\def\h{h}
\def\mh{m_{\h}}
\def\mm{\mu^+\mu^-}
\def\mt{m_t}

\def\mw{M_W}
\def\mz{M_Z}
\def\gsim{\agt}

\begin{document}

\preprint{
\font\fortssbx=cmssbx10 scaled \magstep2
\hbox to \hsize{
\hfill$\vcenter{\hbox{\bf MADPH-96-963}
             \hbox{\bf IUHET-346}
             \hbox{\bf UCD-97-01}
             \hbox{January 1997}}$ }
}

\title{\vspace*{.75in}
Precision $W$-boson and top-quark mass determinations\\
at a muon collider}

\author{V. Barger$^a$, M.S.~Berger$^b$, J.F.~Gunion$^c$, and T.~Han$^c$}

\address{
$^a$Physics Department, University of Wisconsin, Madison, WI 53706, USA\\
$^b$Physics Department, Indiana University, Bloomington, IN 47405, USA\\
$^c$Physics Department, University of California, Davis, CA 95616, USA}

\maketitle

\thispagestyle{empty}

\begin{abstract}

Precise determinations of the masses  of the $W$ boson and of the top quark
could stringently test the radiative structure of the Standard Model (SM) or
provide evidence for new physics. We analyze the excellent prospects at
a muon collider for measuring $M_W$ and $m_t$
in the $W^+W^-$ and $t\bar t$ threshold regions. With an integrated
luminosity of 10 (100)~fb$^{-1}$,
the $W$-boson mass could be measured to
a precision of 20 (6)~MeV, and the top-quark mass
to a precision of 200 (70)~MeV, provided that
theoretical and  experimental systematics are understood.
A measurement of $\Delta m_t=200$~MeV for fixed $M_W$
would constrain a 100 GeV SM Higgs mass within about $\pm 2$~GeV,
while $\Delta M_W=6$~MeV for fixed $m_t$
would constrain $m_h$ to about $\pm 10$~GeV.

\end{abstract}

\newpage

\section{Introduction}

Muon colliders offer a wide range of opportunities for exploring physics
within and beyond the Standard Model
(SM)~\cite{mupmumi,saus,montauk,sanfran,feas}.
An important potential application of these machines is the precision
measurement of  particle masses, widths and couplings.
Because the muon mass is much larger than the electron mass, initial state
radiation from muons is substantially reduced compared to that from
electrons and higher precision is possible in measuring cross sections
in threshold regions.
In this report, we estimate the accuracy with which the $W$ and $t$ masses
can be determined from $\wp\wm$ and $t\anti t$ threshold measurements
at a muon collider. We find that a muon collider with high luminosity may
achieve greater accuracy for $\mw$ and $\mt$ than any other accelerator.

The $W$ and $Z$ masses are related by the equation
\begin{equation}
\mw = \mz
\left [1-{{\pi\alpha}\over {\sqrt{2}G_\mu \mw^2(1-\delta r)}}\right ]^{1/2},
\label{eq:WZmasses}
\end{equation}
where $\delta r$ represents loop effects \cite{loops}. In the SM,
$\delta r$ depends quadratically on the top-quark
mass and logarithmically on the Higgs mass ($\mh$).
In the supersymmetric SM, $\delta r$ may in addition depend on the masses
of light supersymmetric particles, such as the chargino and top squark.
The $W$ and $t$ mass errors should have
a relative precision of
\begin{equation}
\Delta \mw \sim 0.7\times 10^{-2} \Delta m_t
\label{idealratio}
\end{equation}
in order that they lead to equivalent error in testing Eq.~(\ref{eq:WZmasses}).
The present world averages~\cite{demar}
for the $W$ boson mass and the top-quark mass are
\begin{eqnarray}
M_W=80.356 \pm 0.125~{\rm GeV}\;, \quad
m_t=175 \pm 6~{\rm GeV}\;,
\end{eqnarray}
for which $\Delta M_W/\Delta m_t \sim 2\times10^{-2}$.

With high precision measurements of $M_Z$, $M_W$ and $m_t$, the consistency of
the SM loop corrections can be tested and used to infer the Higgs mass
through $\delta r$ in  Eq.~(\ref{eq:WZmasses}).
Figure~1 shows SM predictions of $M_W$ (on-shell
mass definition \cite{bernd}) versus $m_t$  for $m_h = 70$, 100 and 1000~GeV.
With the present $M_W$ and $m_t$ measurements (the data point with
error bars in Fig.~1), it is not yet possible to make a
definitive distinction between the light Higgs ($m_h \sim 100$~GeV) and heavy
Higgs ($m_h \sim 1$~TeV) scenarios. As future precision measurements narrow the
allowed Higgs mass range, the results can be confronted with search limits or
direct measurements of the Higgs boson mass.

\begin{center}
\epsfxsize=4.75in\hspace{0in}\epsffile{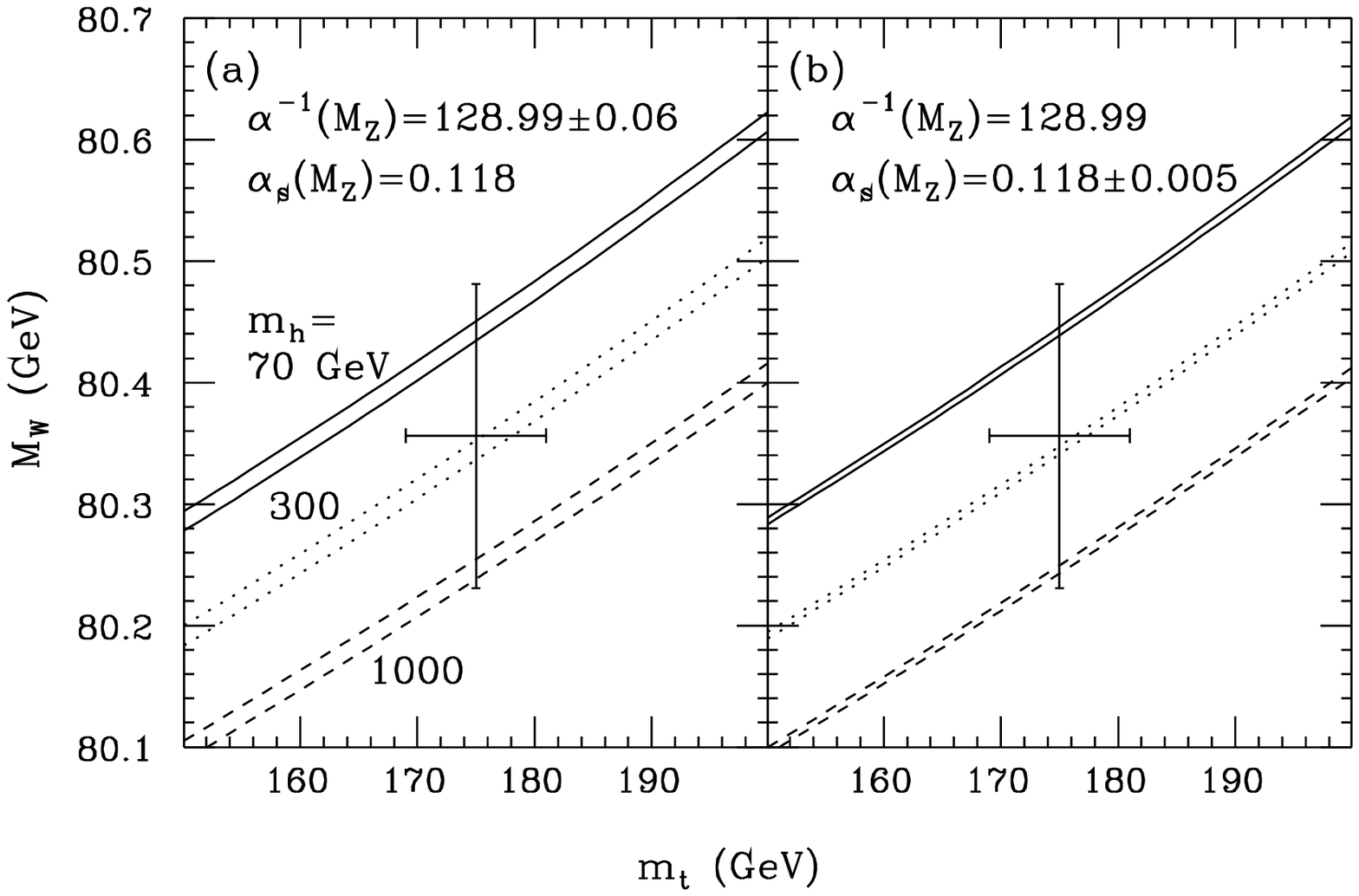}

\medskip
\parbox{5.5in}{\footnotesize\sf Fig.~1: Correlation between $M_W$ and $m_t$
in the SM with QCD and electroweak corrections
\protect\cite{bernd} for $m_h=70,300$ and 1000~GeV. The data point
and error bars are for
$M_W=80.356 \pm 0.125$~GeV and $m_t = 175 \pm 6$~GeV.
The widths of the bands indicate the uncertainties
(a) in $\alpha(M_Z)$ and (b) in $\alpha_s(M_Z)$.}
\end{center}
\medskip

The widths of the bands in Fig.~1 are due to the present uncertainties
in the electromagnetic fine structure constant \cite{alpha1}
and in the strong coupling constant \cite{alphas}
\begin{equation}
\alpha ^{-1}(M_Z)=128.99\pm 0.06\ , \quad \alpha_s(M_Z)=0.118\pm 0.005\;.
\label{Dieters}
\end{equation}
These errors in $\alpha ^{-1}(M_Z)$ and in $\alpha_s(M_Z)$
translate into uncertainties on $M_W$ of order 20 MeV and 4 MeV
respectively. Thus, an improvement in the uncertainty
in the fine structure constant at the $Z$ mass scale will
be needed to fully utilize a very accurate determination of $\mw$. The
uncertainty in $\alpha^{-1}(M_Z)$  is due to measurement errors in
the cross section for $e^+e^- \rightarrow \rm hadrons$.
The determination of $\alpha ^{-1}(M_Z)$ given in Eq.~(\ref{Dieters})
made use of the perturbative QCD cross section
in the energy region above 5 GeV in order to effectively eliminate
that source of the error.
Half of the remaining uncertainty in $\alpha^{-1}(M_Z)$
in Eq.~(\ref{Dieters}) is due to the 1 to
2.5~GeV region and the other half from the 2.5--5~GeV region.
Calculations of  $\alpha^{-1}(M_Z)$
have been made by other authors \cite{alpha2} without
the perturbative QCD parameterization of the data in the region
5--35~GeV.
The cross section in the 1--5~GeV energy range for $e^+e^-\to\rm hadrons$,
now measured to 10\% accuracy,
could possibly be improved to 1\% accuracy at the Beijing,
Frascati or Novosibirsk machines \cite{alpha1}. Measurements
at this level of precision
would translate into an overall error on $\alpha^{-1}(M_Z)$ of order
$\pm0.03$ \cite{DZ-private}. A 1\% determination of $\alpha_s(M_Z)$
from future high energy experiments and lattice calculations is also
anticipated \cite{alphas}.

Because of the importance of testing the radiative structure of the theory, it
is appropriate to consider what improvements in precision measurements can be
made at other colliders \cite{bauretal}.
The planned upgrades of the Tevatron collider
will lead to improved $M_W$ and $m_t$
determinations through measurements of the $e\nu$ transverse mass and other
techniques. With the Main Injector (MI, operational in 1999) and
possible TeV33 upgrade, the anticipated precisions are \cite{bauretal}
\begin{eqnarray}
\Delta M_W &=& 50 \pm 20{\rm\ MeV}, \quad
\Delta m_t = 4\rm\ GeV \qquad (2\ fb^{-1},\ MI) \\
\Delta M_W &=& 20{\rm\ MeV}, \quad
\Delta m_t = 2\rm\ GeV \qquad (10\ fb^{-1},\ TeV33) \,.
\end{eqnarray}
At the large hadron collider (LHC), the expected accuracy for low-luminosity
running is \cite{bauretal,lhcmw}
\begin{equation}
\Delta M_W = 15 {\rm\ MeV},  \qquad
\Delta m_t = 2\rm\ GeV  \qquad (10\rm\ fb^{-1},\ LHC) \,.\label{lhc}
\end{equation}
Running the LHC at a higher luminosity (100 fb$^{-1}$/year) is
actually less effective for precision mass measurements
due to the large background problem \cite{lhcmw}.

Experiments are currently underway at LEP2 that will determine $M_W$ by two
methods \cite{lepii}.
The first is the measurement of the cross section at 161~GeV,
just above the $2M_W$ threshold, to determine the mass via the SM prediction
for the cross section. A precision of
\begin{equation}
\Delta M_W = 144\rm\ MeV \qquad (100\ pb^{-1},\ LEP2)
\label{lep2like}
\end{equation}
is anticipated.
This number assumes an integrated luminosity of 25~pb$^{-1}$ in
each of the 4 experiments. The second method is the reconstruction of the $W$
mass in
$W\to jj$ decay modes from $W^+W^-\to q \bar{q} q\bar{q}$ and $W^+W^-\to q\bar
q\ell\nu$ final states at $\sqrt s = 175$~GeV where the cross section is
much larger. The achievable precision in this case is
\begin{equation}
\Delta M_W = 34\rm\ MeV \qquad (2\ fb^{-1},\ LEP2) \,.
\end{equation}
At a linear $e^+e^-$ collider (NLC), the anticipated precisions
for $M_W$ measurement \cite{nlcmw} with $\sqrt s=500$ GeV and for $m_t$
measurement \cite{nlcmt} at the threshold $\sqrt s \sim 2m_t$ are
\begin{equation}
\Delta M_W = 20 {\rm\ MeV}, \qquad
\Delta m_t = 0.2\rm\ GeV  \qquad (50\rm\ fb^{-1},\ NLC) \,.\label{nlc}
\end{equation}
where the $\mw$ error is that for mass reconstruction in the $q\anti q$
decay mode. If the NLC energy is lowered to $\rts=161\gev$ for a $\wp\wm$
threshold determination of $\mw$, then scaling the statistical error
of Eq.~(\ref{lep2like}) implies $\Delta\mw\sim 20\mev$ ($6\mev$)
for an integrated luminosity of
$L=5\fbi$ ($50\fbi$), where the former is a rough estimate
of the yearly luminosity that would result at $\rts=161\gev$ for
an interaction region designed for $L=50\fbi$ per year at $\rts=500\gev$.
The actual NLC errors would be larger, however, since systematic
uncertainty in the central beam energy value as well as the beam
energy spread would cause significant deterioration; detailed estimates
are not available. In this paper, we show that
such problems are minimal at a muon collider: an accuracy
for $\mw$ near the statistical level is possible and the accuracy
on $\mt$ would also be higher than at the NLC. For an integrated luminosity of
$50\fbi$,
we find that $\Delta \mw\sim 9\mev$ and $\Delta\mt\sim 100\mev$
can be achieved at a muon collider provided experimental systematic errors in
cross section
ratios related to detection efficiencies and certain
theoretical systematic errors are sufficiently small.

The prospects for measuring the $W$ boson mass at a muon collider were examined
previously by Dawson \cite{dawson},
who concentrated on the case where
only 100~pb$^{-1}$ luminosity was available for the measurement. 
In this paper, we assume that the muon collider ring is 
\newpage
\noindent
optimized for the $W$ threshold
study and that up to 100~fb$^{-1}$ is available.\footnote{Since the storage
rings would comprise a modest fraction
of the overall collider costs \cite{palmer}, it should be possible to have
separate rings optimized for the threshold energies and then high luminosities
can be realized.}
In this situation one must confront
the systematic errors that may dominate over the statistical ones.
We find rough agreement with the scaled statistical error in
Ref.~\cite{dawson}. We consider the systematic errors and how to minimize them
in Section~II.

The outline of the remainder of the paper follows.
In Section II, we examine the accuracy
with which $\mw$ can be determined using cross section measurements
at a $\mu^+ \mu^-$ collider near the $W^+W^-$
threshold. Determining $\mt$ via measurements near the $t\anti t$
threshold is discussed in Section~III.
In Section IV, we summarize our results and emphasize the
constraints on the SM Higgs boson mass from the precision
$M_W$ and $m_t$ measurements.

\section{$M_W$ Measurement at the $\mu^+\mu^-\to W^+W^-$ Threshold}

In lepton collider measurements of $M_W$ through reconstruction of
$W \rightarrow q \bar q$ decays, the hadronic calorimeter resolution
determines the achievable precision. The
$WW \rightarrow q\bar q \ell \nu$ final state is preferable to
$WW \rightarrow 4$ jets to avoid potential uncertainties from
the rescattering of quarks originating from different $W$-bosons.

The alternative method of $M_W$ determination based on accurate
cross section measurement near the threshold is
insensitive to the final state $M_W$ reconstruction. A muon collider
is particularly
well suited to the threshold measurement because the energy of the beam has
a very narrow spread. Thus, it is this approach that we
concentrate on here.

The off-shell $W^+W^-$ cross section that comprises the signal is
\begin{equation}
\sigma (s) = \int _0^sds_1\int _0^{(\sqrt{s}-\sqrt{s_1})^2}ds_2
\rho(s_1)\rho(s_2)\sigma_0(s,s_1,s_2)\left [1+\delta_C(s,s_1,s_2)\right ]\;,
\end{equation}
where $\sigma_0$ is the Born cross section given in Ref.~\cite{mnw} and
\begin{equation}
\rho(s)={1\over \pi}{\Gamma _W\over M_W}
{s\over {(s-M_W^2)^2+s^2\Gamma_W^2/M_W^2}}\;.
\end{equation}
The form for the Coulomb correction $\delta _C$
can be found in Refs.~\cite{coulomb}.
Initial state radiation (ISR) must also be included in the cross section
calculation. Since the radiative effects are smaller for muons than for
electrons, the signal cross section is slightly higher at a $\mm$ collider.
The predicted signal at a muon collider is plotted in Fig.~2 for several values
of $M_W$.

\begin{center}
\epsfxsize=4.75in\hspace{0in}\epsffile{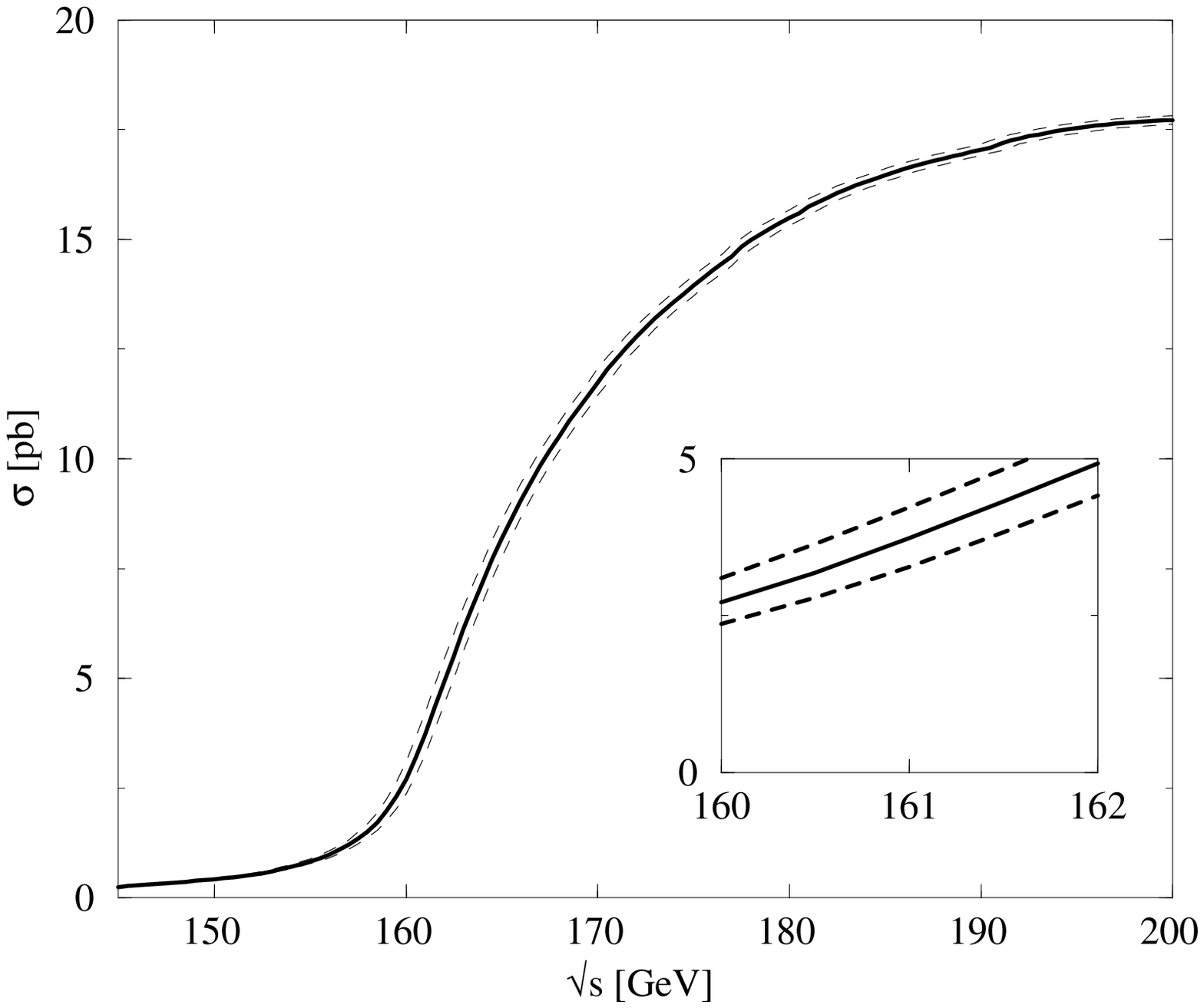}

\medskip
\parbox{5.5in}{\footnotesize\sf Fig.~2: The cross section for $\mm \to W^+W^-$
in the threshold region for $\mw=80.3$~GeV (solid) and $\mw=80.1,80.5$~GeV
(dashed).
The inlaid graph shows the region of the threshold curve where the
statistical sensitivity to $\mw$ is maximized.
Effects of ISR have been included.}
\end{center}
\bigskip

The threshold cross section is most
sensitive to $\mw$ just above $\sqrt s = 2\mw$,
but a tradeoff exists between
maximizing the signal rate and the sensitivity of the cross section to $\mw$.
Detailed analysis \cite{lepii} shows that
if the background level is small and systematic uncertainties in efficiencies
are not important, then the optimal measurement of $\mw$
is obtained by collecting data at a single energy
$$\sqrt{s} \sim 2\mw + 0.5  \gev\ \sim\ 161  \gev ,$$
where the threshold cross section is sharply rising.

For a LEP2 measurement with 100~pb$^{-1}$ of integrated luminosity
the background and systematic uncertainties are, in fact, sufficiently small
that the error for $\mw$ will be limited by the statistical uncertainty
of the measurement at $\rts=161\gev$.  But, at a muon
collider or electron collider at high luminosity, systematic errors
arising from uncertainties in the background level and the detection/triggering
efficiencies will be dominant unless
some of the luminosity is devoted to measuring
the level of the background (which automatically includes somewhat similar
efficiencies) at an energy below the $W^+W^-$ threshold.
Then, assuming that efficiencies for the background and $\wp\wm$ signal are
sufficiently well understood that systematic uncertainties
effectively cancel in the ratio
of the above-threshold to the below-threshold rates, a very accurate $\mw$
determination becomes possible.

The dominant background derives from $e^+e^- \to
(Z/\gamma) (Z/\gamma)$ which is essentially
energy independent~\cite{lepii} below
180~GeV. For our present analysis we model the background as energy
independent, and accordingly assume that one measurement at an energy in the
range 140 to 150~GeV suffices to determine the background.

Thus, we analyze our ability to determine the $W$ mass via
just two measurements: one at center of mass energy $\sqrt{s}=161\gev$, just
above threshold, and one at $\sqrt{s}=150\gev$. The signal is not entirely
negligible at the lower energy (especially in the $q\overline{q}\ell \nu$ and
$\ell \nu \ell \nu$ modes) due to off-shell $W$-decay contributions,
but a two-parameter fit for $M_W$ and the
(constant) background can be made.
The optimal $M_W$ measurement is obtained by expending about two-thirds
of the luminosity at $\sqrt{s}=161\gev$ and  one-third
at $\sqrt{s}=150\gev$.
We assume the signal detection efficiencies 
(not including branching fractions)
of 55\%, 47\% and 60\% for the decay modes
$WW\to q\overline{q}q\overline{q}, q\overline{q}\ell \nu, \ell \nu \ell \nu$
respectively along with the background cross-section with cuts from
Ref.~\cite{lepii}.

Our joint determination of the signal (and hence the measurement of
$M_W$) and background levels is shown in Fig.~3 for the mode
$WW\to q\overline{q}q\overline{q}$.
These results, for an integrated luminosity of 100~fb$^{-1}$, indicate that a
determination of
$M_W$ to a precision of 9~MeV is possible in this final state, with the
underlying background measured to 1\% accuracy. Note that this
1\% characterizes the level at which the systematic efficiency uncertainties
must be under control in the ratio of the 161 and 150 GeV cross section
measurements. The $\pm 9\mev$ uncertainty for $M_W$ is equivalent
to about a 0.5\% measurement of the signal cross section as
apparent in Fig.~2, where the inset shows
that a 200~MeV shift in $\mw$ results in about at 10\% shift in the cross
section.

\begin{center}
\epsfxsize=4.75in\hspace{0in}\epsffile{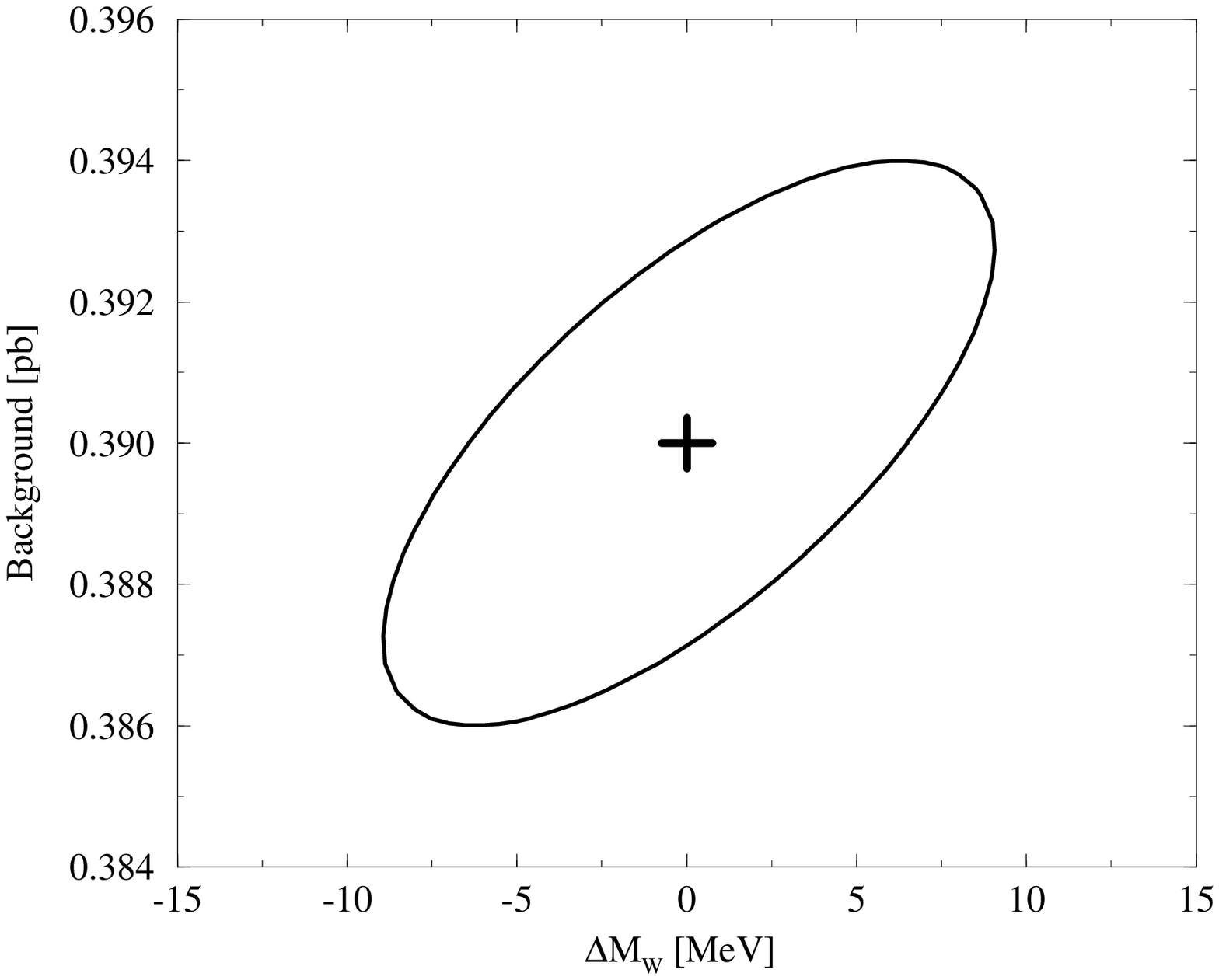}

\medskip
\parbox{5.5in}{\footnotesize\sf Fig.~3: A sample $\Delta \chi^2=1.0$ contour
for the background and signal measurement in the $W^+W^-$
threshold region for the final state $q\overline{q}q\overline{q}$
with an integrated luminosity of $100~{\rm fb}^{-1}$.
Effects of ISR have been included.
A central mass value $M_W=80.356$~GeV is assumed.}
\end{center}
\bigskip

Table~I lists the achievable $M_W$ precision in the various $W$ decay modes for
$100~{\rm fb}^{-1}$ of integrated luminosity.
Combining the three modes, an overall precision of
\begin{equation}
\Delta M_W = 6 \rm\ MeV
\end{equation}
should be achievable.
The above analysis assumes the predicted SM width $\Gamma_W$. An uncertainty  
in $\Gamma_W$ may translate into an uncertainty in $M_W$ since $\Gamma_W$  
can affect the cross section at the threshold. 
Quantitatively the relation at $\sqrt s = 161$~GeV
is\footnote{This is similar to the case at LEP2 \cite{lepii}, 
even though the cross section dependence on $M_W$ and $\Gamma_W$ 
at a muon collider is stronger.} $\Delta M_W \simeq 0.17\Delta\Gamma_W$.
The current experimental error $\Delta\Gamma_W \simeq 60$~MeV \cite{datab} 
thus translates into an uncertainty of 
$\Delta M_W\sim 10$~MeV. With a $10\fbi$ integrated luminosity
at the Tevatron upgrade, $\Delta\Gamma_W \simeq 20$~MeV 
can be achieved \cite{tev2000}, which translates into an
uncertainty $\Delta M_W\sim 3$~MeV. It is interesting to note 
that there is essentially no dependence of the cross section on 
$\Gamma_W$ near $\sqrt s = 162$~GeV, so that the uncertainty
due to $\Delta\Gamma_W$ can be minimized by performing the 
measurements at this energy without degrading the
$M_W$ determination.

\bigskip
\vbox{
\begin{center}
\parbox{5.5in}{\footnotesize Table I: Signal and background cross sections for
$\mu^+\mu^-\to W^+W^-$ and the achievable precision in $M_W$  with
100~fb$^{-1}$ luminosity. A central mass value $M_W=80.356$~GeV is assumed.}
\medskip

\def\arraystretch{1.25}
\tabcolsep=1em
\begin{tabular}{|c|c|c|c|}
\hline
&$q\overline{q}q\overline{q}$& $q\overline{q}\ell \nu$& $\ell \nu \ell \nu$
\\ \hline
Signal [pb] at $\sqrt{s}=161$~GeV & 0.97 & 0.77 & 0.25\\ \hline
Signal [pb] at $\sqrt{s}=150$~GeV & 0.11 & 0.086 & 0.028\\ \hline
Background [pb]& 0.39 & 0.03 & 0.01\\ \hline
$\Delta M_W$ [MeV]& 9 & 8 & 14\\ \hline
\end{tabular}
\end{center}
} 
\bigskip

Let us return to the issue of theoretical and experimental
systematic uncertainties.
On the theoretical side, there are uncertainties
in $M_W$ from mass definition schemes and the renormalization scale
which may be on the order of a few MeV \cite{bernd}.
Therefore, careful theoretical consideration
is required to extract the precise (\eg\ $\overline{\rm MS}$)
$M_W$ value and test loop effects.
Higher order corrections to the signal cross section
remain to be evaluated, but such calculations are not a serious obstacle.

The largest systematic
effects, however, may be those associated with the
background treatment and systematic errors in the detection/triggering
efficiencies. First, a more refined treatment of the energy dependence
of the background may prove to be necessary.
Theoretical calculations can be used to input the
energy dependence and checked via
measurements made at more than one sub-threshold energy. Excellent
accuracy on the energy dependence should be possible.
The biggest
uncertainty is likely to arise from lack of knowledge of the efficiencies.
In particular, the background and the signal are
somewhat different in that the background has different leptonic modes
and different percentages of jet-jet final states relative to leptonic final
states. It will be crucial that the efficiency for
background final states relative to that for signal final states
be understood to better than the 1\% level.
Alternatively, the technique of determining $\mw$
by measuring the $\wp\wm$ signal well
above threshold, $\rts\gsim 200\gev$, and taking the ratio to the
$\rts=161\gev$ measurement could also be considered --- since the
final states involved are the same, efficiencies for detection/triggering
may cancel to the needed degree of accuracy.

Uncertainty in $M_W$ due to uncertainty in the beam energy is roughly
given by $\Delta M_W \simeq \Delta E_{beam}$ \cite{lepii}. At a muon 
collider $\Delta E_{beam} < 10^{-5} E_{beam}$ is achievable \cite{palmer},
implying $\Delta M_W \leq 0.8\mev$. Beam energy smearing will also 
have negligible impact on $M_W$ so long as the Gaussian width 
is known and much less than $\Gamma_W$. Finally, the relative luminosity 
at $\sqrt s=150\gev$ and $161\gev$ must be known to better than 0.5\%
for the systematic error from this source to yield 
$\Delta M_W < 6 \mev$.

%
\section{Top-quark Mass Measurement at the
$\mu^+\mu^-\to \lowercase{t \bar t}$ Threshold}

There is very rich physics associated with the $t\bar t$ threshold, including
the determination of $m_t$, $\Gamma_t$ ($|V_{tb}|$), $\alpha_s$, and
possibly $m_h$ \cite{kuhn}.
A precise value of the top-quark mass $m_t$ could prove to be
very valuable in theoretical studies. For example, if a particle desert exists
up to the GUT
scale,  we will want to extrapolate from low-energy to the
grand unified scale to probe in a detailed way the physics at the unification
scale. The
top-quark mass (and its Yukawa coupling) are crucially important
since they determine to a large extent the evolution of all the
other Yukawa couplings, including flavor mixings.
If the top-quark Yukawa coupling is determined by an infrared quasi-fixed
point \cite{bbo}, very small changes in $m_t$ translate into very large changes
in the renormalized values of many other parameters in the theory.

Fadin and Khoze first
demonstrated that the top-quark threshold cross section is
calculable since the large
top-quark mass puts one in the perturbative regime of QCD,
and the large top-quark width effectively screens nonperturbative effects
in the final state \cite{fk}.
Such studies have since been performed by several
groups \cite{feigenbaum,kwong,sp,jht,bagliesi,sfhmn,immo,fms}.
There are two equivalent ways to obtain the total cross section near
threshold by solving for a three-point Green's function in either
coordinate \cite{sp} or momentum space \cite{jht}.
Here we solve Schr\"{o}dinger's equation
in coordinate space
\begin{equation}
\left [-{{\Delta }\over m_t}+V(r)-\left (E+i{\Gamma _{\Theta}\over 2}\right )
\right ]G({\bf x};E) = \delta^3({\bf x})\;,
\end{equation}
where $\Gamma _{\Theta}$ is the (running) toponium width,
and $E=\sqrt{s}-2m_t$.
The potential $V(r)$ is given at small $r$ by two-loop perturbative QCD and
for large $r$ by a fit to quarkonia spectra.
In our analysis we make use of the Wisconsin potential~\cite{wiscp} that
interpolates these regimes.
However, the short range part of the potential
alone determines the physics at the top threshold.
The cross section is proportional to ${\rm Im}\; G({\bf x}=0;E)$
with \cite{sp,fms}
\begin{equation}
\sigma _{t\bar{t}}={{96\pi ^2\alpha ^2}\over s^2}
\left \{1-{{16\alpha_s}\over 3\pi}\right \}
[(Q_eQ_t+v_ev_t\chi)^2+(a_e^2v_t^2\chi^2)]
{\rm Im}\; G({\bf x}=0;E=\sqrt{s}-2m_t)\;,
\end{equation}
where $\chi = s/(s-M_Z^2)$.
The  cross section depends on the strong gauge coupling
$\alpha _s(M_Z)$ through the potential $V(r)$.

Figure~4 shows the calculated threshold curve for $\mu^+\mu^-$
or $e^+e^- \to t \bar{t}$
including the effects of ISR
for a top-quark mass of 175~GeV.
The initial state radiation causes a reduction of the
cross section as well as a smearing of the small resonance peak.
The effect is less severe for a muon collider (long-dashed)
than that for an $e^+e^-$ collider (short-dashed) due to the
heavier muon mass.

\begin{center}
\epsfxsize=3.9in
\hspace*{0in}
\epsffile{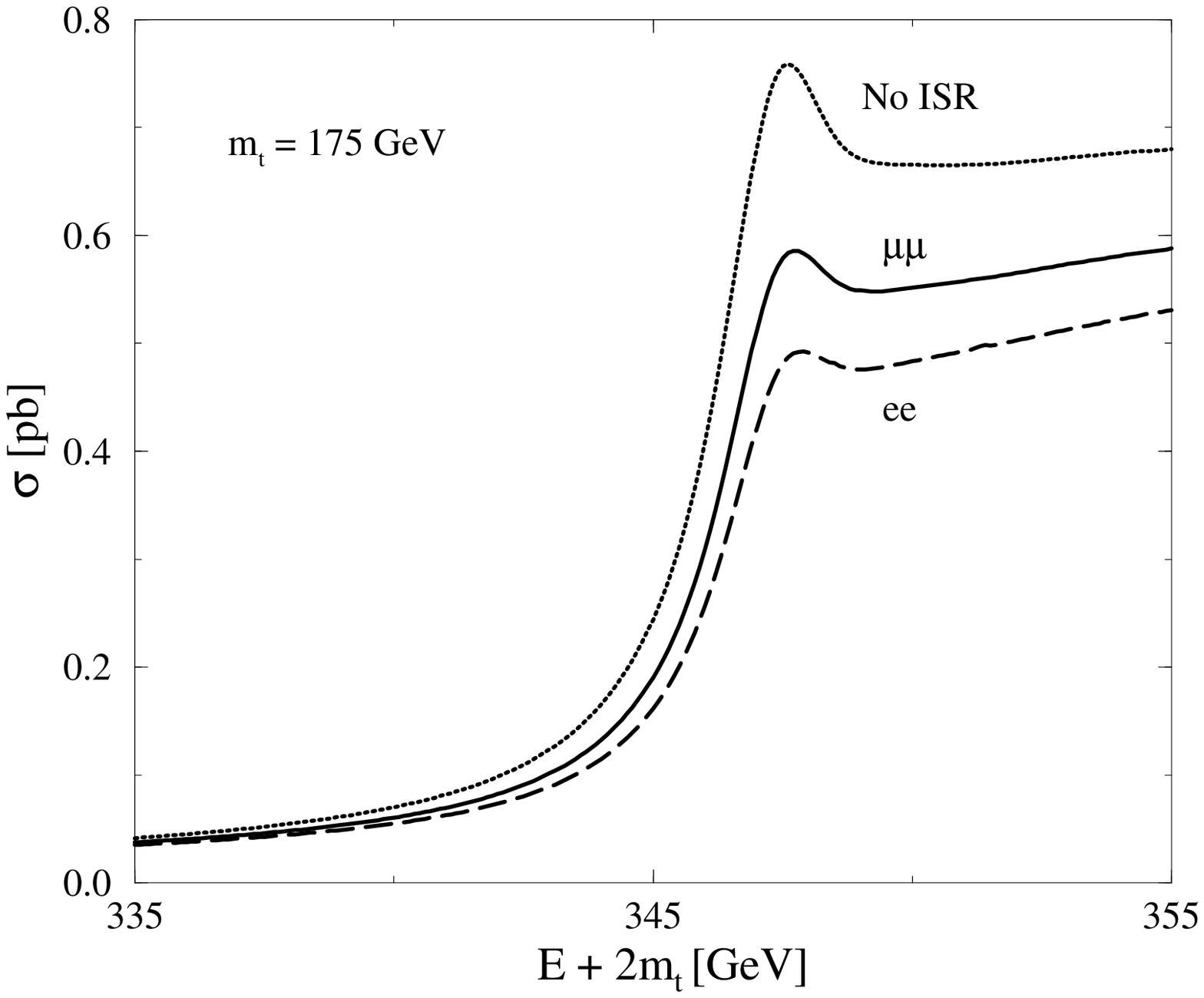}

\parbox{5.5in}{\footnotesize\sf Fig.~4:
The cross section for $t \bar{t} $ production at a lepton collider
in the threshold region, for $m_t=175$ GeV and
$\alpha _s(M_Z)=0.12$. The results for $\mu\mu$ and $ee$ colliders include the
effects of ISR (but no beam smearing), and the top curve does not.}
\end{center}
\medskip

The beam energy spread is the major experimental problem in precision
measurements of the
top threshold region at an $e^+e^-$ collider. Ref.~\cite{bagliesi}
demonstrated the effects of beam smearing for some proposed $e^+e^-$ machine
designs, and argued that a narrow beam was essential for studying the
$t\anti t$ threshold region. A high resolution
determination of the $e^+e^-$ collider energy profile is desirable in order to
be able to deconvolute the smearing of the threshold curve.

For a muon collider a
measurement of the beam profile is unnecessary, since  a very narrow
beam is a natural characteristic.
The rms deviation $\sigma $ in $\sqrt{s}$ is given by \cite{bbgh1,bbgh2}
\begin{equation}
\sigma = (250~{\rm MeV})\left({R\over 0.1\%}\right)\left({\sqrt s\over {\rm
360\ GeV}}\right) \;,
\end{equation}
where $R$ is the rms deviation of the Gaussian beam profile.
With $R\alt 0.1\%$ the resolution $\sigma$ is of the same
order as the measurement one hopes to make in the top mass.
For $t\overline{t}$ studies the exact shape of the beam is not
important if $R\alt 0.1\%$. We take
$R=0.1\%$ here; the results are not improved significantly with better
resolution.

Changing the value of the strong coupling
constant $\alpha _s(M_Z)$ influences
the threshold region. Large values lead to
tighter binding and the peak shifts to lower values of $\sqrt{s}$.
Weaker coupling also smooths out the threshold peak.
These effects are illustrated in Fig.~5.

\begin{center}
\epsfxsize=3.8in
\hspace*{0in}
\epsffile{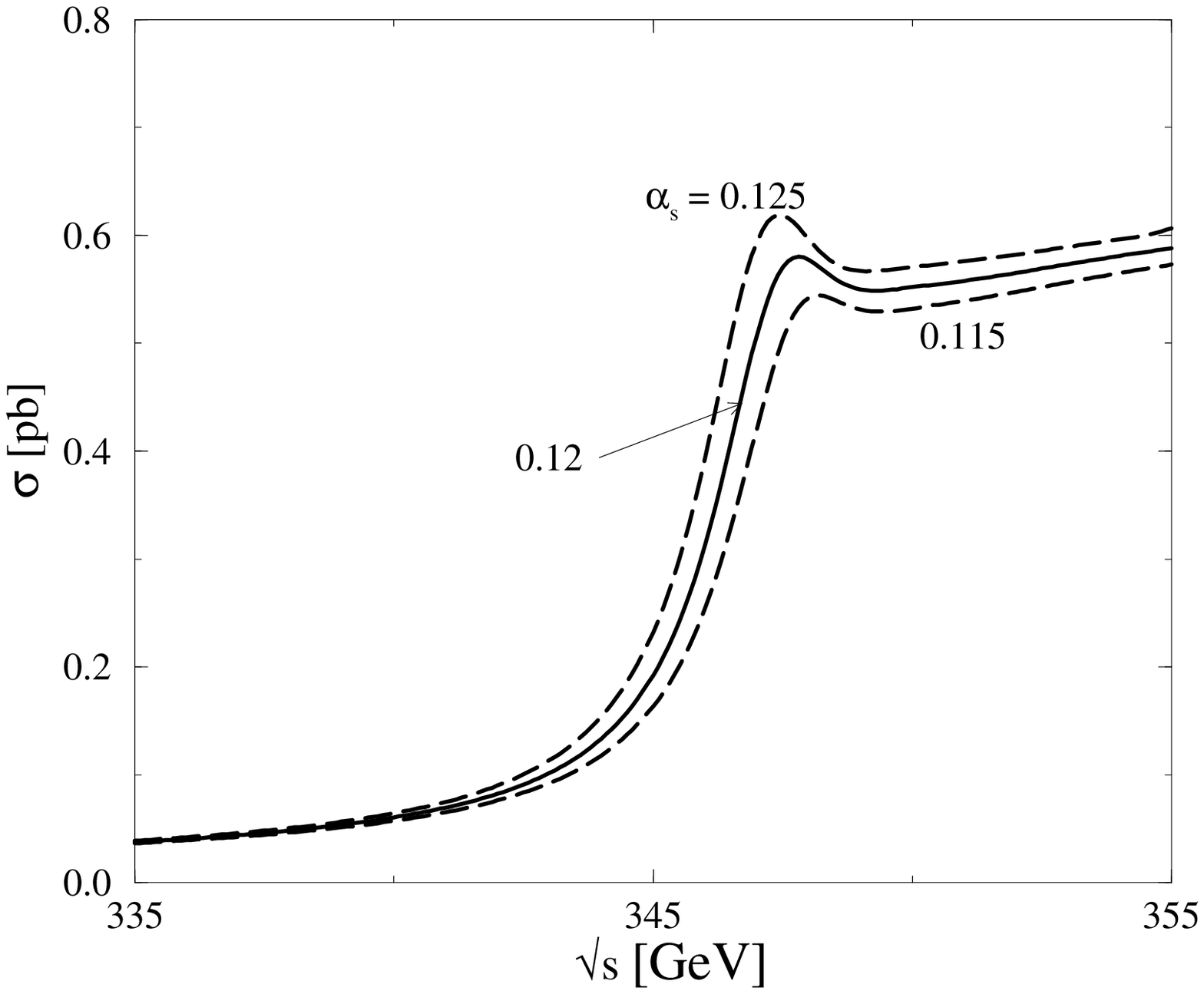}

\parbox{5.5in}{\footnotesize\sf Fig.~5:
The cross section for $\mu^+\mu^- \to t \bar{t}$
production in the threshold region, for $m_t=175$ GeV and
$\alpha_s(M_Z)=0.12$ (solid) and 0.115, 0.125 (dashes).
Effects of ISR and beam smearing are included.}
\end{center}
\medskip

To assess the precision of parameter determinations from
cross section measurements, we generate hypothetical sample data,
shown in Fig.~6, assuming that 10~fb$^{-1}$ integrated luminosity
is used to measure the cross section at each energy in
1~GeV intervals. Since the top threshold curve
depends on other quantities like $\alpha_s(M_Z)$, one must do a full scan to
determine the shape of the curve and its overall normalization.
To generate the ten
data points in Fig.~6 we use nominal values of $m_t=175$ GeV
and $\alpha_s(M_Z)=0.12$.
Following Ref.~\cite{fms} we assume a 29\% detection efficiency for
$W\to q\bar q$, including the decay branching fraction. 
The data points can then be fit to theoretical
predictions
for different values of $m_t$ and $\alpha_s(M_Z)$; the
likelihood fit that is obtained is shown
as the $\Delta \chi ^2$ contour plot in Fig.~7.
The inner and outer curves are the $\Delta \chi^2=1.0$ (68.3\%) and $4.0$
(95.4\%)
confidence levels respectively for the full 100~fb$^{-1}$ integrated
luminosity.
Projecting the $\Delta \chi^2=1.0$ ellipse on the $m_t$ axis,
the top-quark mass
can be determined to within $\Delta\mt\sim 70\mev$,
provided systematics are under control. (Systematic error issues
will be discussed later.)
A top-quark mass of 175 GeV can be measured to about 200 MeV at
$90\%$ confidence level with 10~fb$^{-1}$ luminosity.
This is about a factor of 1.7 better in $\Delta m_t$
than the same measurement at an
$e^+e^-$ machine when realistic beam effects are included \cite{fms}.

\begin{center}
\epsfxsize=3.8in
\hspace*{0in}
\epsffile{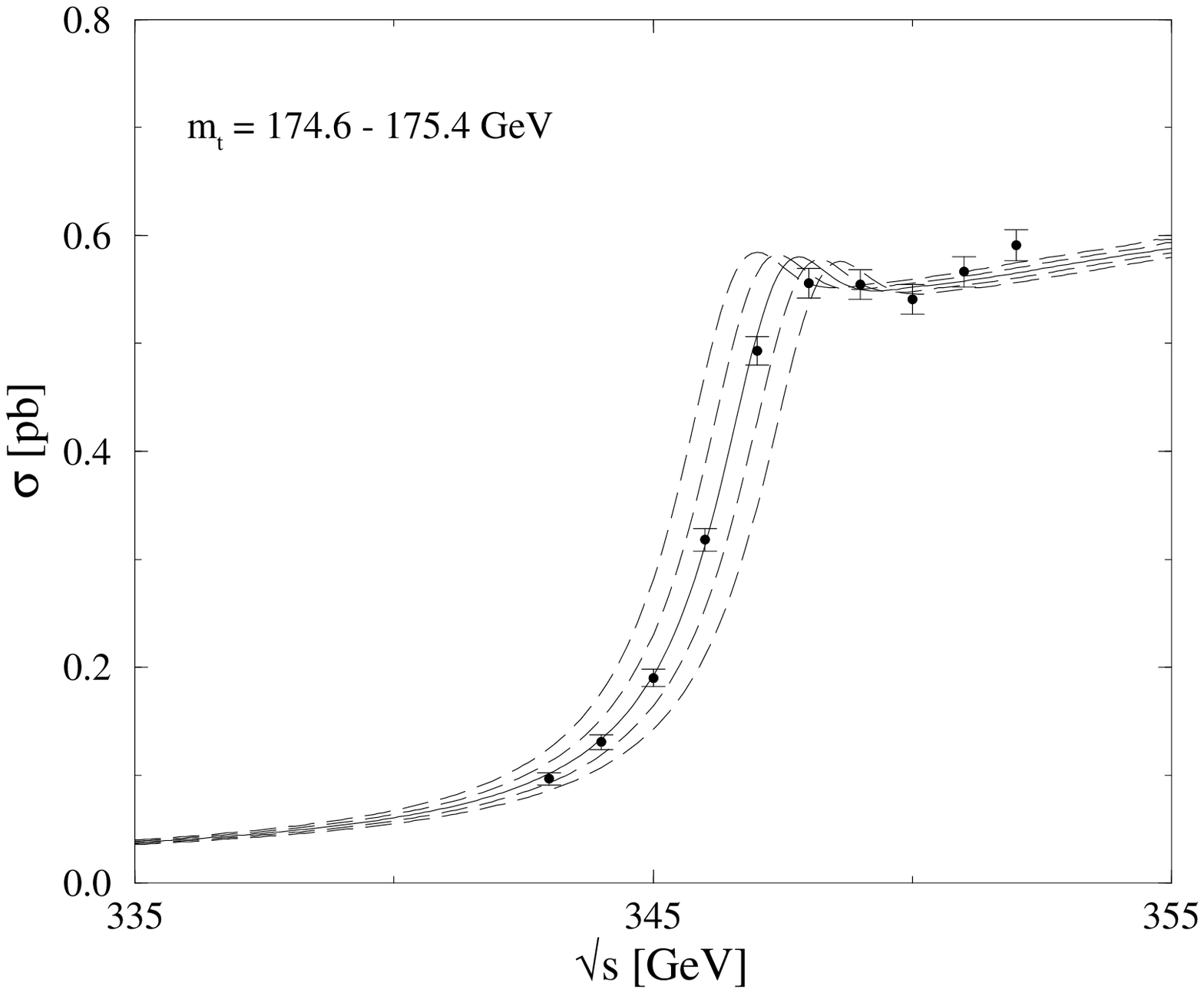}
\medskip

\parbox{5.5in}{\footnotesize\sf Fig.~6: Sample data for $\mu^+ \mu^-
\rightarrow t \bar{t} $ obtained assuming a scan over
the threshold region devoting 10~fb$^{-1}$ luminosity to each data point.
A detection efficiency of 29\% has been assumed \cite{fms}
in obtaining
the error bars. The threshold curves correspond to shifts in $m_t$ of 200~MeV
increments. Effects of ISR and beam smearing have been included,
and the strong coupling $\alpha_s(M_Z)$ is taken to be 0.12.}
\end{center}
\bigskip

Since the exchange of a light Higgs boson can affect the threshold shape,
a scan of the
threshold cross section can in principle yield some information about the Higgs
mass and its Yukawa coupling to the top quark.
Figure~8 shows the dependence of the threshold curve on the Higgs mass, $m_h$.
The effect of the Higgs boson vertex correction can be
obtained \cite{higgs} by including a Yukawa interaction in the QCD potential,
\begin{equation}
V_h(r) = -{{\sqrt{2}G_F}\over {4\pi r}}m_t^2e^{-m_hr}\;,
\label{higgspot}
\end{equation}
which effectively results in multiplying the resulting cross section by a small
energy-independent correction factor.\footnote{Eq.~(\ref{higgspot}) assumes
the SM Higgs to $t\anti t$ coupling.
In the case of the minimal supersymmetric model, the couplings of the
lightest Higgs boson $h$ become very similar to those of the SM
Higgs boson in the large $m_A$ limit (where $A$ is the CP-odd neutral
Higgs boson) \cite{hunters}.}
However, it may be difficult to disentangle such a Higgs
effect from two-loop QCD effects, which are not yet fully
calculated~\cite{hoang}.

\begin{center}
\epsfxsize=3.9in
\hspace*{0in}
\epsffile{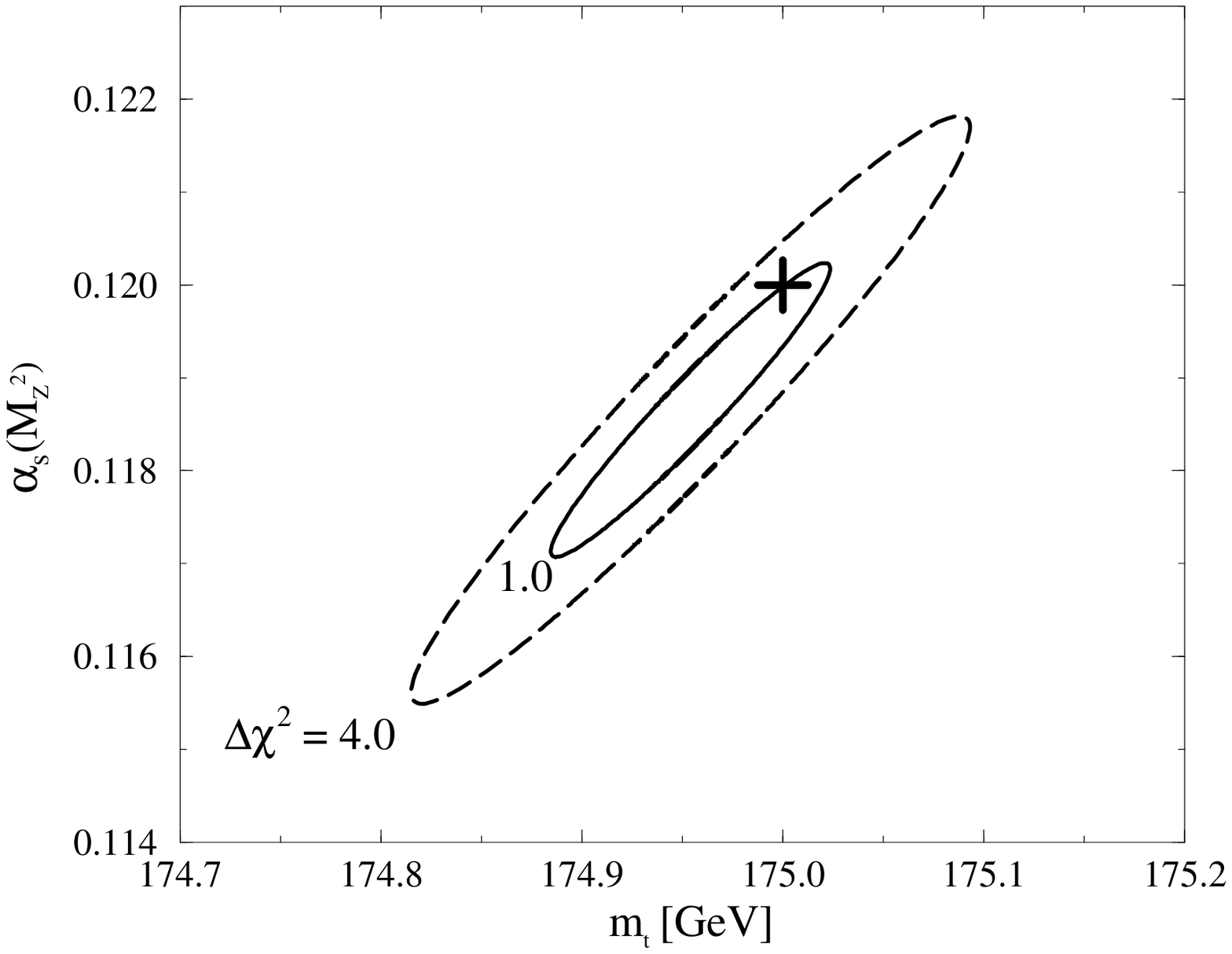}

\parbox{5.5in}{\footnotesize\sf Fig.~7: The $\Delta \chi ^2=1.0$ and
$\Delta \chi ^2=4.0$ confidence
limits for the sample data shown in Fig.~6. The ``+''
marks the input values from which the data were generated.}
\end{center}
\medskip

In addition to the Higgs Yukawa potential effect, there is an additional
$s$-channel Higgs contribution \cite{bbgh1,bbgh2} to the
cross section at a muon collider since the muon has a larger Yukawa coupling
than does the
electron; however, the $s$-channel contribution is much smaller than the usual
photon and $Z$ exchanges considered here.

Changing the top-quark width from its value in the Standard Model also
affects the threshold shape. The width can be parameterized in terms of the
CKM element $|V_{tb}|$, for which one expects $|V_{tb}|\approx 1$ in the
Standard Model.  A value $|V_{tb}|>1$
would indicate new physics contribution to the top-quark decay, such as
$t\to bH^+$.
The dependence on $|V_{tb}|^2$ is shown in Fig.~9.
A narrower top quark (smaller $|V_{tb}|$) results in a more prominent
1S peak in the cross section.

\begin{center}
\epsfxsize=3.9in
\hspace*{0in}
\epsffile{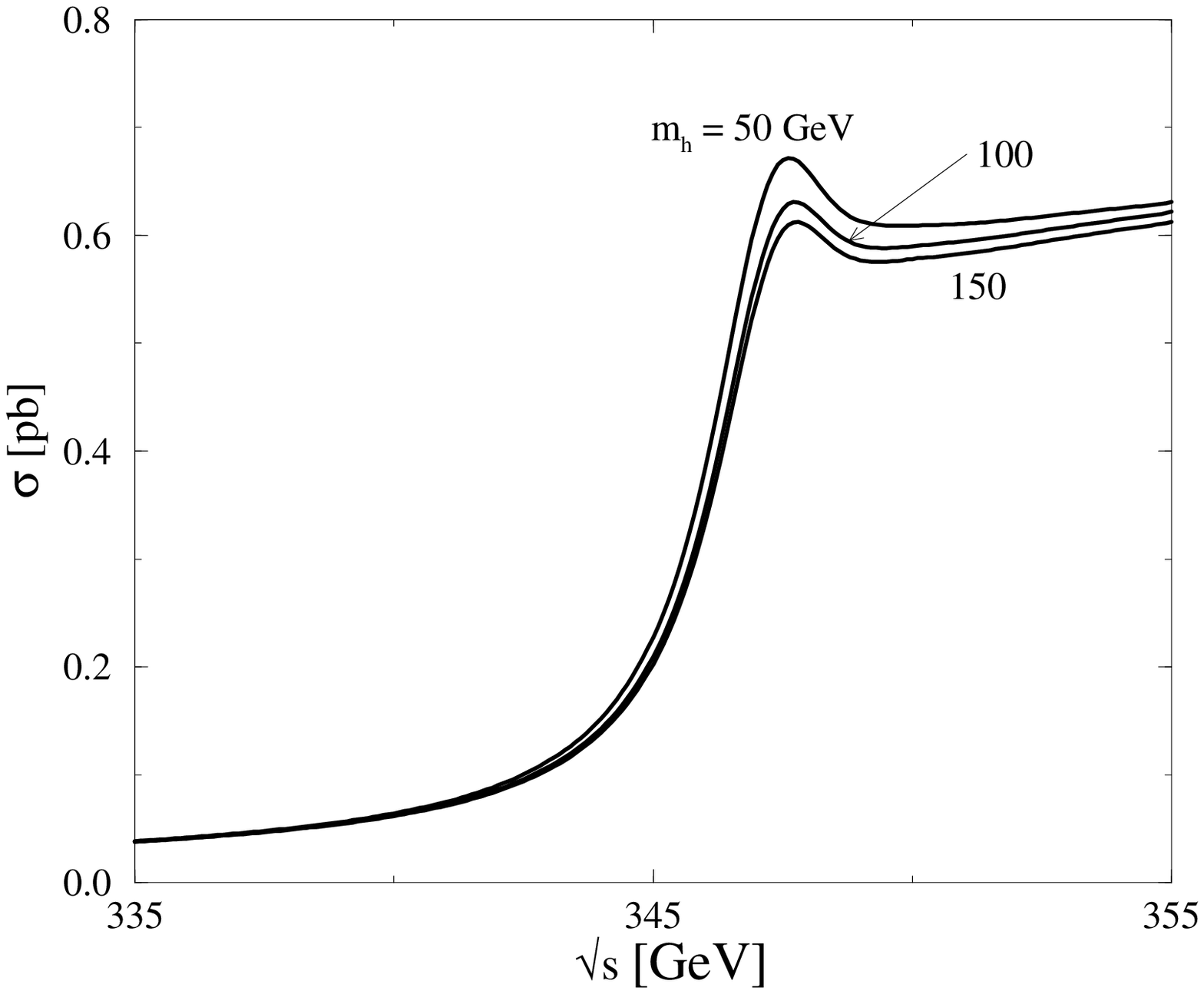}

\parbox{5.5in}{\footnotesize\sf Fig.~8: The dependence of the threshold region
on the Higgs mass, for $m_h=50, 100, 150$~GeV. Effects of
ISR and beam smearing have been included,
and we have assumed $m_t=175$~GeV and $\alpha_s(M_Z)=0.12$.}
\end{center}
\medskip

\begin{center}
\epsfxsize=3.9in
\hspace*{0in}
\epsffile{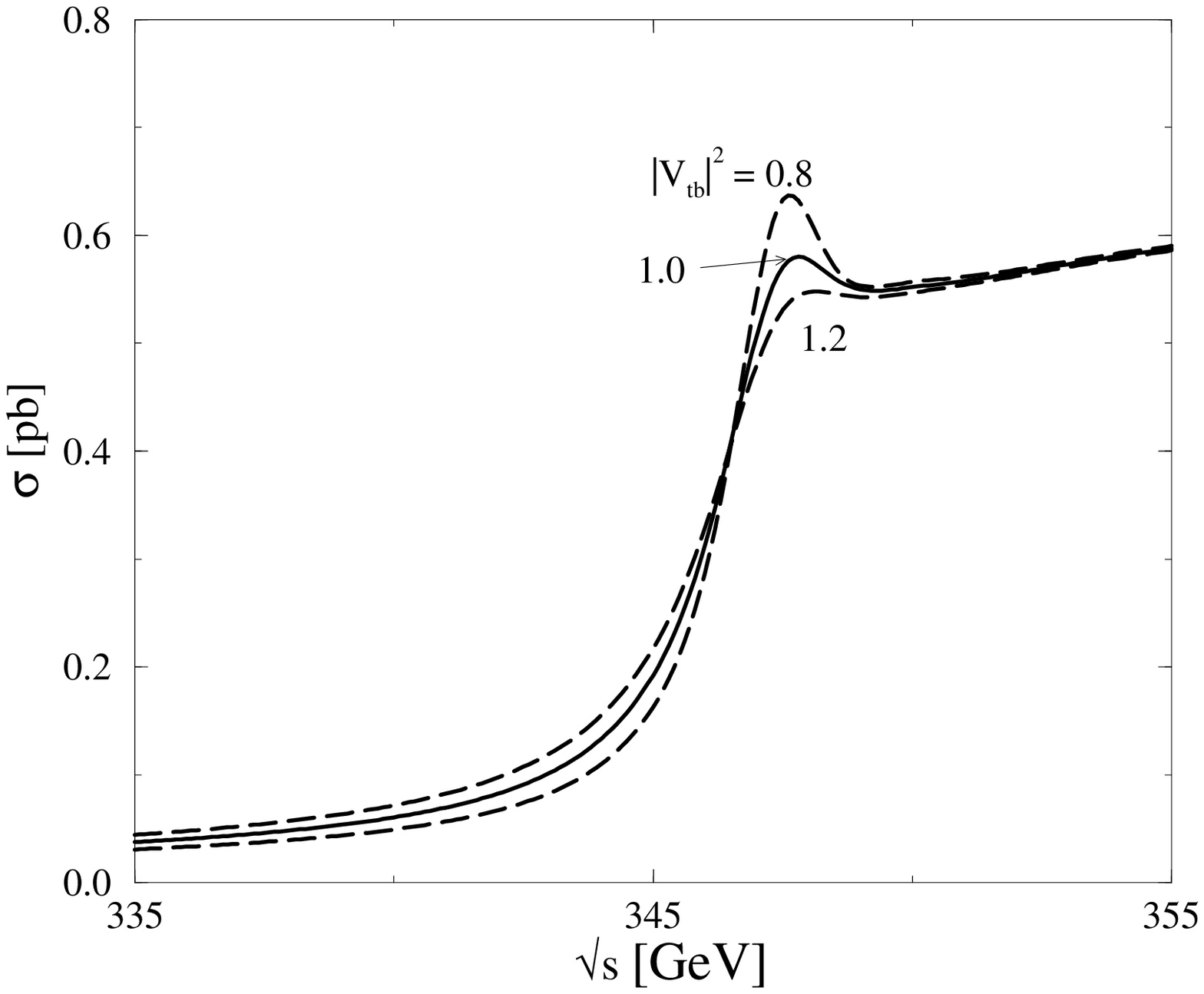}

\parbox{5.5in}{\footnotesize\sf Fig.~9: The dependence of the
threshold region for $\mu^+ \mu^- \rightarrow t \bar{t}$ production
on the $|V_{tb}|^2=0.8,1.0,1.2$. Effects of ISR and beam smearing have
been included, and we have assumed $m_t=175$~GeV and $\alpha_s(M_Z)=0.120$.}
\end{center}
\bigskip

It should be possible to experimentally distinguish the various effects on the
$t\bar{t}$ threshold shape. In Figure 10 we show the dependence of four
quantities
\begin{eqnarray}
{\rm a)}\ &&[\sigma({\rm peak})-\sigma(340)]/\sigma(340)\;, \\ \nonumber
{\rm b)}\ &&[\sigma(350)-\sigma(340)]/\sigma(340)\;, \\ \nonumber
{\rm c)}\ &&[\sigma({\rm peak})-\sigma(350)]/\sigma(350)\;, \\ \nonumber
{\rm d)}\ &&\sqrt{s}_{\rm peak}\;. \nonumber
\end{eqnarray}
on the parameters $m_t$, $\alpha_s$, $m_h$ and $|V_{tb}|$.
Here $\sqrt{s}_{\rm peak}$ is the c.m.\ energy of the peak in the cross section
and $\sqrt{s}=340$ and 350 GeV are energies above and below this peak. These
four quantities show different dependencies on the four parameters;
consequently,
detailed fits to  threshold data should determine the parameters.
In Fig.~10, the central values of $m_t=175$~GeV, $\alpha_s=0.12$
and $|V_{tb}|^2=1.0$ were
chosen and then one parameter was varied to make the corresponding curve.
The $m_h$ curves show the effect of including a Higgs Yukawa contribution
to the potential.

\begin{center}
\epsfxsize=5.9in
\hspace*{0in}
\epsffile{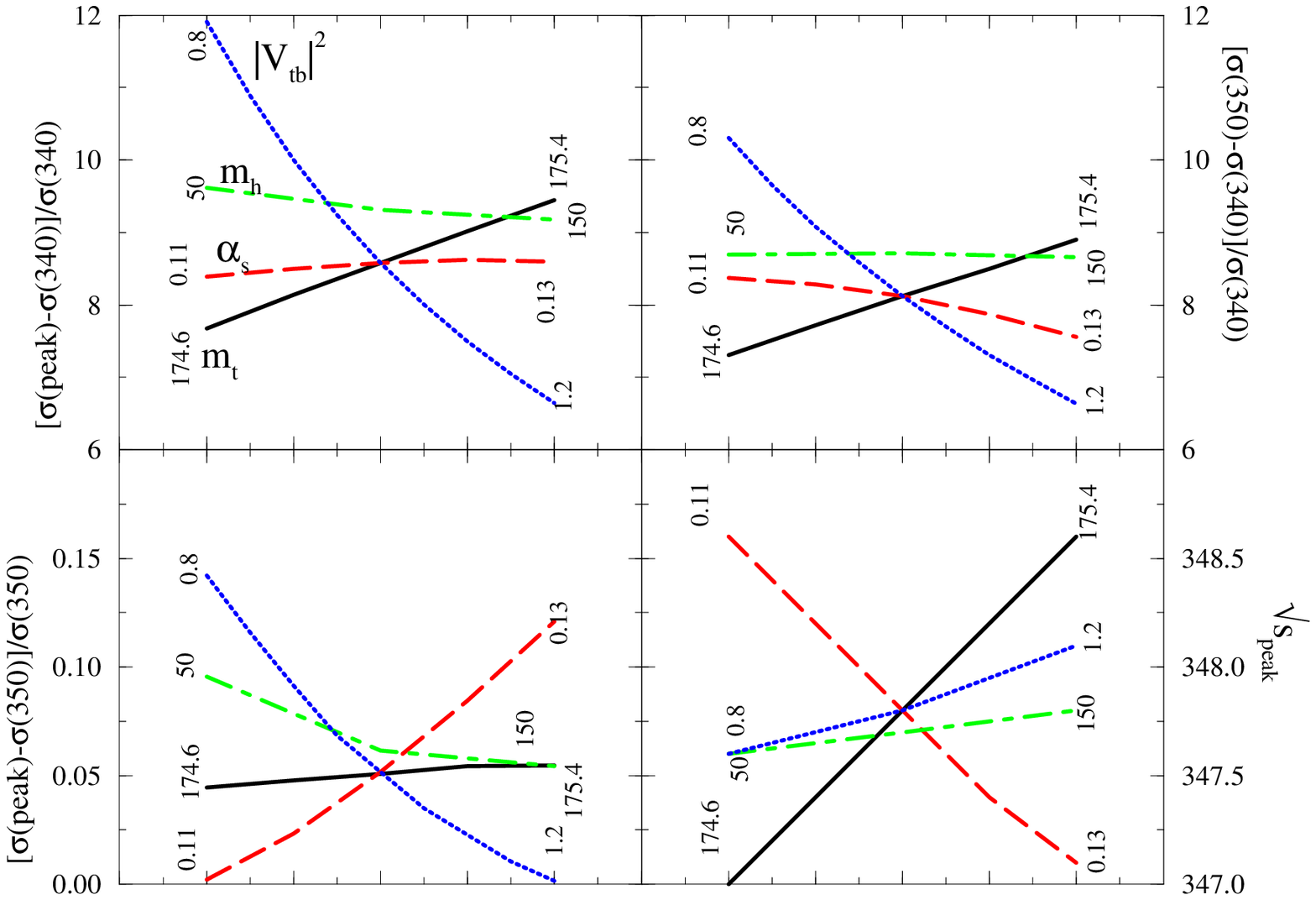}
\vspace*{-0.5in}

{\footnotesize\sf Fig.~10: The dependence of various measurable
ratios on $m_t$, $\alpha_s$, $m_h$ and $|V_{tb}|^2$.
The Higgs Yukawa potential of Eq.~(\ref{higgspot})
is only included in the $m_h$ curves.}
\end{center}
\medskip

Additional information can be obtained by measuring the top-quark
momentum from a reconstruction of the decay \cite{fms},
providing further constraints on
$\alpha_s$ and $|V_{tb}|$ (the dependence on $m_t$ is small).

We now consider a variety of systematic uncertainties/issues.
\begin{itemize}
\item
QCD measurements at future colliders and
lattice calculations will presumably determine
$\alpha_s(M_Z)$ to 1\% accuracy (e.g.\ $\pm 0.001$) \cite{alphas}
by the time muon colliders are constructed so the uncertainty in
$\alpha_s$ will likely be similar to
the precision obtainable at a $\mu^+\mu^-$ and/or $e^+e^-$
collider with 100~fb$^{-1}$ integrated luminosity.
If the luminosity available for the threshold measurement is
significantly less than 100~fb$^{-1}$,
one can regard the value of $\alpha_s(M_Z)$ coming from
other sources as an input, and thereby improve the
top-quark mass determination.
\item
There is some theoretical ambiguity in the
mass definition of the top quark. The theoretical
uncertainty on the quark pole mass
due to QCD confinement effects is of  order
$\Lambda_{QCD}$, {\it i.e.},  a few hundred MeV \cite{sw}.
In the $\overline{\rm MS}$ scheme of quark mass definition, the theoretical
uncertainty is better controlled.

\item
Systematic errors in experimental efficiencies are not a significant
problem for the $t\anti t$ threshold determination of $\mt$.  This can
be seen from Fig.~6, which shows that a 200 MeV shift in $\mt$ corresponds
to nearly a 10\% shift in the cross section
on the steeply rising part of the
threshold scan, whereas it results in almost no change in $\sigma$
once $\rts$ is above the peak by a few GeV.  Not only will efficiencies
be known to much better than 10\%, but also systematic uncertainties
will cancel to a high level of accuracy in the ratio of
the cross section measured above the peak to measurements
on the steeply-rising part of the threshold curve.
\item
As Fig.~8 shows,
it will be important to know the Higgs mass and the $h t\anti t$
coupling strength in order to eliminate this source of systematic
uncertainty when extracting other quantities.

\end{itemize}

The measurements described in this section can be performed at either an
$e^+e^-$ or a $\mu^+\mu^-$ collider.
The errors for $\mt$ that we have found for the muon collider are smaller than
those previously obtained in studies at the NLC electron
collider primarily because
the smearing of the threshold region by the energy spread
of the beam is much less, and secondarily due to the fact that the reduced
amount of initial state radiation makes the cross section somewhat larger.

\section{Conclusion}

A muon collider offers an unparalleled opportunity for precision $W$ and
top-quark mass measurements in the respective threshold regions.
Table II compares the precision achievable for $M_W$ and $m_t$
at present and future colliders.

\medskip
\begin{center}
\parbox{5.5in}{\footnotesize Table II: Comparison for the
achievable precision in $M_W$ and $m_t$ measurement at
different future colliders.}
\medskip

\def\arraystretch{1.25}
\tabcolsep=1em
\begin{tabular}{|c|c|c|c|c|c|c|c|c|}
\hline
            & \multicolumn{2}{c|}{\quad LEP2}&
\multicolumn{2}{c|}{\quad Tevatron}& LHC& NLC&
\multicolumn{2}{c|}{\quad $\mu^+\mu^-$} \\ \hline
$\cal L$ (fb$^{-1}$)& 0.1& 2& 2& 10& 10& 50& 10& 100\\ \hline
$\Delta M_W$ (MeV)& 144& 34& 35& 20& 15& 20& 20& 6\\ \hline
$\Delta m_t$ (GeV)&  --& --& 4&  2& 2& 0.2& 0.2& 0.07 \\ \hline
\end{tabular}
\end{center}
\bigskip

We summarize our main results as follows:

\begin{itemize}

\item At the $W$ threshold,
the optimum strategy is to expend about
2/3 of the luminosity at $\sqrt s = 161$~GeV, just above $2M_W$,
and about 1/3 at $\sqrt s = 150$~GeV to measure the background
and normalize efficiencies.
With 10 (100)~${\rm fb}^{-1}$ of integrated luminosity
at a muon collider, $M_W$ could be measured to a precision
of 20 (6)~MeV, provided that
the theoretical cross sections for the $W^+W^-$ signal are evaluated
to the $\alt {\cal O}(1\%)$ level and that no irreducible
systematic (in particular, experimental errors for
cross section ratios) remain at this level.

\item
With an integrated luminosity of 10 (100)~${\rm fb}^{-1}$,
the top-quark mass can be measured to 200 (70)~MeV,
using a 10-point scan over the threshold region, in 1~GeV intervals,
to measure the shape predicted by the QCD potential.
In the $t \bar{t} $ threshold study, differences of cross sections at energies
below, at, and above the resonance peak, along with the location of the
resonance peak, have different dependencies on the parameters
$m_t$, $\alpha_s$, $m_h$ and $|V_{tb}|^2$
and should allow their determination.
To utilize the highest precision measurements achievable at the statistical
level, theoretical uncertainties
and other systematics need to be under control. We are confident
that uncertainty in $\alpha_s$ will not be a factor and we have
noted that ratios of above-peak
measurements to measurements on the steeply rising part of the
threshold cross section will eliminate many experimental systematics
related to uncertainties in efficiencies.

\begin{center}
\epsfxsize=4.75in\hspace{0in}\epsffile{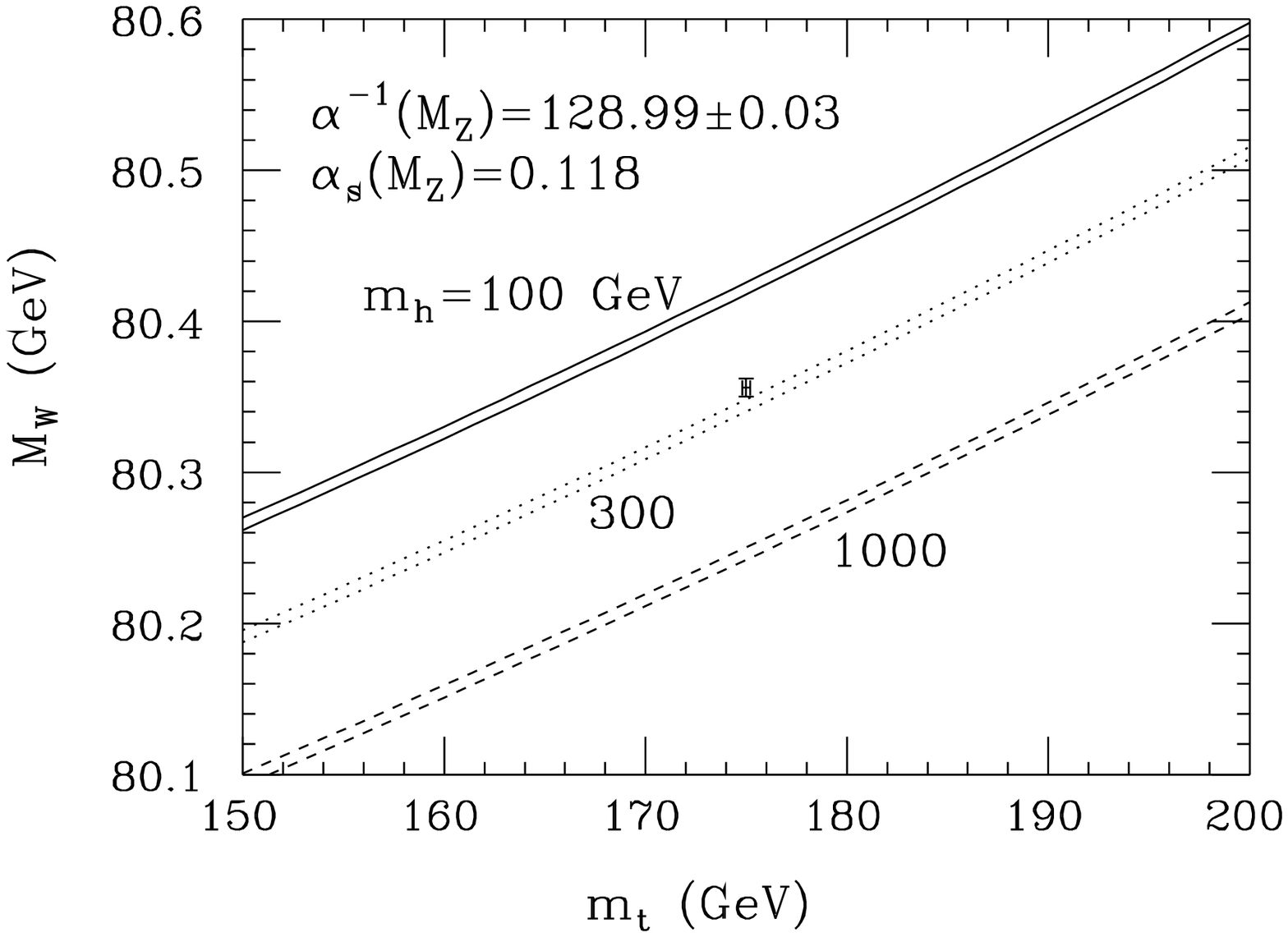}

\medskip
\parbox{5.5in}{\footnotesize\sf Fig.~11:
Correlation between $\protect M_W$ and $\protect m_t$
in the SM with QCD and electroweak corrections
for $\protect m_h=100, 300$ and 1000~GeV. The data point
and error bars illustrate the possible accuracy for
the indirect $m_h$ determination assuming
$M_W=80.356\pm 0.006$ GeV and $m_t=175\pm 0.2$ GeV.
The widths of the bands indicate the uncertainty
in $\protect \alpha(M_Z)$.}
\end{center}
\medskip

\item
The combination of the measurements of
the masses $M_Z$, $M_W$ and $m_t$ to such high precision has dramatic
implications
for the indirect prediction of the mass of the Higgs boson and for other
sources of physics beyond the Standard Model. This is illustrated in
Fig.~11. Assuming the current central values
of $M_W$, $m_t$,  $\alpha(M_Z)$ and $\alpha_s(M_Z)$, and that $L=10\fbi$
($100\fbi$) is devoted to the measurement of $\mt$ ($\mw$),
the mass of the SM Higgs
boson would be determined to be 260 GeV with an error of about $\pm 5$~GeV from
$\Delta m_t=200$~MeV at a fixed $M_W$,
and about $\pm 20$~GeV from $\Delta M_W=6$~MeV at a fixed $m_t$.
For $m_h=100$~GeV, the corresponding values would be
$\pm 2$~GeV  and $\pm 10$~GeV, respectively.
More generally, the $\Delta m_h$ value scales roughly like $m_h$.

\end{itemize}

Concerning the indirect determination of $m_h$ from the
radiative correction relations, there is no
need to devote more than $10\fbi$ of luminosity
to determining $\mt$; indeed, the ideal ratio of Eq.~(\ref{idealratio})
would be reached for just $L\sim 0.6\fbi$ (yielding
$\Delta\mt\sim 900\mev$) if $\Delta\mw\sim 6\mev$.  The low luminosity needed
at $\rts\sim 2\mt$ could probably be accumulated without difficulty
using a ring optimized for $\rts\sim 2\mw$.

An accuracy of $\Delta M_W \sim 6\mev$ achievable at a muon
collider would approach
the precision level of the current $M_Z$
measurements. It will test the consistency of the Standard 
Model at the multi-loop level, whatever the Higgs mass
value is, or probe physics beyond the SM.
A low-energy muon collider program that explores 
$\wp\wm$ and $t\bar t$ threshold prodcution,
$s$-channel Higgs production \cite{bbgh1,bbgh2}, and $Zh$ 
threshold production \cite{Zh} could have enormous impact  
on SM physics and beyond.

\section*{Acknowledgments}

We thank B.~Kniehl for providing us with a program to evaluate SM
radiative corrections, and
D.~Zeppenfeld and S.~Willenbrock
for discussions about the determination of $\alpha$
and the top-quark pole mass. We also thank M. Peskin for
a discussion on top-quark threshold physics.
This work was supported in part by the U.S. Department of Energy
under Grants No. DE-FG02-95ER40896, No.~DE-FG03-91ER40674 and
No.~DE-FG02-91ER40661.
Further support was provided
by the University of Wisconsin Research
Committee with funds granted by the Wisconsin Alumni Research
Foundation, and by the Davis Institute for High Energy Physics.

\end{document}